\documentclass[aps,pra,reprint,superscriptaddress]{revtex4-1}
\usepackage{graphicx}
\usepackage{amsmath}
\usepackage{amssymb}
\usepackage[colorlinks,linkcolor=blue,
            citecolor=blue,
            hyperindex,
            pdfstartview=FitH,
            plainpages=false]
            {hyperref}
\usepackage{soul}

\newcommand {\bscco}{Bi$_2$Sr$_2$CaCu$_2$O$_{8+\delta}$}
\newcommand {\bsco}{Bi$_2$Sr$_{2-x}$La$_x$CuO$_6$}
\newcommand {\Pbbsco}{Bi$_{1.76}$Pb$_{0.35}$Sr$_{1.89}$CuO$_{6+\delta}$}

\begin{document}
\title{Resolving unoccupied electronic states with laser ARPES in bismuth-based cuprate superconductors}

\author{Tristan L. Miller}
\affiliation{Materials Sciences Division, Lawrence Berkeley National Laboratory, Berkeley, California 94720, USA}
\affiliation{Department of Physics, University of California, Berkeley, California 94720, USA}
\author{Minna \"Arr\"al\"a}
\affiliation{Department of Physics, Tampere University of Technology, PO Box 692, FIN-33101 Tampere, Finland}
\author{Christopher L. Smallwood}
\author{Wentao Zhang}
\affiliation{Materials Sciences Division, Lawrence Berkeley National Laboratory, Berkeley, California 94720, USA}
\affiliation{Department of Physics, University of California, Berkeley, California 94720, USA}
\author{Hasnain Hafiz}
\author{Bernardo Barbiellini}
\affiliation{Department of Physics, Northeastern University, Boston, Massachusetts 02115}
\author{Koshi Kurashima}
\affiliation{Department of Applied Physics, Tohoku University, Sendai 980-8579, Japan}
\author{Tadashi Adachi}
\affiliation{Department of Engineering and Applied Sciences, Sophia University, Tokyo 102-8554, Japan}
\author{Yoji Koike}
\affiliation{Department of Applied Physics, Tohoku University, Sendai 980-8579, Japan}
\author{Hiroshi Eisaki}
\affiliation{Electronics and Photonics Research Institute, National Institute of Advanced Industrial Science and Technology, Tsukuba, Ibaraki 305-8568, Japan}
\author{Matti Lindroos}
\affiliation{Department of Physics, Tampere University of Technology, PO Box 692, FIN-33101 Tampere, Finland}
\author{Arun Bansil}
\affiliation{Department of Physics, Northeastern University, Boston, Massachusetts 02115}
\author{Dung-Hai Lee}
\affiliation{Department of Physics, University of California, Berkeley, California 94720, USA}
\author{Alessandra Lanzara}
\email{alanzara@lbl.gov}
\affiliation{Materials Sciences Division, Lawrence Berkeley National Laboratory, Berkeley, California 94720, USA}
\affiliation{Department of Physics, University of California, Berkeley, California 94720, USA}
\date {\today}

\begin{abstract}

Angle-resolved photoemission spectroscopy (ARPES) is typically used to study only the occupied electronic band structure of a material.  Here we use laser-based ARPES to observe a feature in bismuth-based superconductors that, in contrast, is related to the unoccupied states.  Specifically, we observe a dispersive suppression of intensity cutting across the valence band, which, when compared with relativistic one-step calculations, can be traced to two final-state gaps in the bands 6 eV above the Fermi level.  This finding opens up possibilities to bring the ultra-high momentum resolution of existing laser-ARPES instruments to the unoccupied electron states.  For cases where the final-state gap is not the object of study, we find that its effects can be made to vanish under certain experimental conditions.

\end{abstract}

\maketitle

%Introduction

Angle-resolved photoemission spectroscopy (ARPES) is a powerful experimental probe that has been used extensively to image the occupied electronic states of materials in an energy- and momentum-resolved manner\cite{Hufner2003,Damascelli2003}. Since it is based on Einstein's photoelectric effect, it cannot directly probe a material's unoccupied electronic states, but it has nevertheless provided signatures of gaps in the unoccupied states\cite{Dietz1978,Courths1989,Strocov1998,Strocov2012}.  According to the one-step model\cite{Pendry1976}, some electrons, after absorbing photons, may have energies which lie between two unoccupied bands. We call the space between the unoccupied bands a final-state gap, although this gap may be confined to a limited momentum range, and disperse within that range.  The photoemission intensity of these electrons is suppressed, but not suppressed completely, due to the finite widths of the final states.  These finite widths represent the small chance that electrons interacting with the medium, primarily through electron-hole pair creation and plasmonic interaction, will have energy within the final-state gap. Typically, this final-state effect in ARPES is not used to measure unoccupied states, which are instead mapped by inverse photoemission\cite{Yoshida2013} or very-low-energy electron diffraction\cite{Lindroos1986,Lindroos1987,Lindroos1987a,Strocov1998,Strocov2003}.

Here we show that laser-based ARPES\cite{Koralek2007,Liu2008,Kiss2008}, under certain conditions, can be used to map final-state gaps in the electronic states of a material.  This method provides the following advantages with respect to standard synchrotron-based ARPES: (a) improved momentum resolution and greater bulk sensitivity, due to the lower photon energy range available in laser-ARPES (6--7 eV)\cite{Koralek2007}, and (b) access to unoccupied electron states closer to the Fermi level.

Data are shown for cuprate superconductors \bscco{} (Bi2212) and \bsco{} (La-Bi2201) along the $\Gamma$--Y direction of the Brillouin zone, using $\sim$6 eV laser ARPES.  In these measurements, a final-state gap can be seen as a line of suppressed intensity that disperses in momentum.  When the final-state gap crosses the photoemitted valence band, it creates a distortion 100--140 meV below the Fermi level, depending on the photon energy.  A second distortion is seen at 20--50 meV, indicating a second final-state gap. These measurements are found to match the calculations of a fully relativistic one-step model, and they demonstrate the power of existing laser-ARPES instruments to map unoccupied electronic states in a momentum-resolved manner.

%Experimental method
Single crystals of nearly optimally doped Bi2212 ($T_c$ = 91 K), La-Bi2201 ($T_c$ = 33 K), and overdoped \Pbbsco{} (Pb-Bi2201, $T_c$ $\sim$ 5 K) were prepared by the traveling solvent floating zone method.  Samples were cleaved \em{in situ}\em{} at pressures less than $5 \times 10^{-11}$ torr, and probed by an ultraviolet laser pulse, tunable around $\sim$6 eV, generated by quadrupling the frequency of a Ti:sapphire laser\cite{Smallwood2012RSI}.  The laser beam is $s$-polarized, about 10$^{\circ}$ from normal incidence, in the same plane as detected electrons.  The energy resolution is $\sim$22 meV and the momentum resolution is $\sim$0.003 \AA$^{-1}$.

%computational method
The computations follow the same method as in Refs. \cite{Arrala2013,Rienks2014}, using a fully relativistic one-step model\cite{Braun1996, Pendry1976} with multiple scattering theory used for both initial and final states, and taking into account effects of the ARPES matrix element on the photointensity\cite{Bansil1999,Lindroos2002,Sahrakorpi2005}.  The surface potential was modeled as a Rundgren-Malmstr\"om barrier\cite{Malmstrom1980}.  Scattering effects were modeled by adding complex energy- and momentum-independent self-energy corrections to the initial- and final-state energies.  Lastly, since the computations do not account for band renormalization of the spectrum due to electron correlation effects, dispersions of the low-energy bands were renormalized by a factor of $Z$ = 0.4, which is within the range of 0.28 to 0.5 found in the literature\cite{Nieminen2012}.  The potential for Bi2212 was computed with the linear augmented-plane-wave method using the WIEN2k package\cite{Blaha2001} in the framework of density functional theory. 
%only include this last sentence if I show any Bi2201 computations
%The potential for Bi2201 was computed by using self consistent electronic structure calculations with Korringa-Kohn-Rostoker method\cite{Kaprzyk1990,Bansil1999}.

%%%%%%%%%%%%%%%%%%%%%%%%%%%%%%%%%
%Fig1
%%%%%%%%%%%%%%%%%%%%%%%%%%%%%%%%%
\begin{figure*}\centering\includegraphics[width=4.75in]{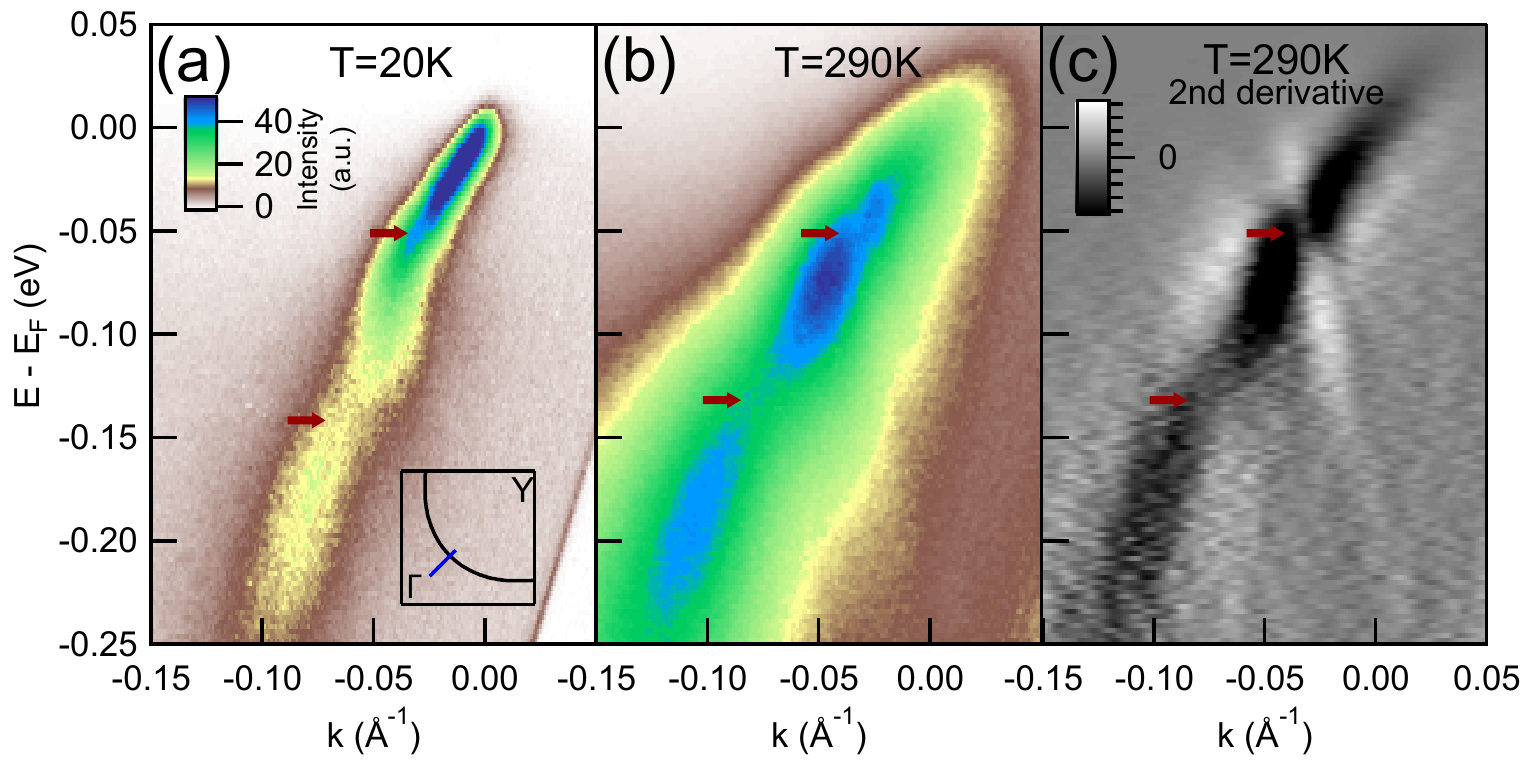}
\caption{(Color online) ARPES intensity maps of optimally doped Bi2212 at 20K (a) and at 290K (b) along the $\Gamma$--Y direction as shown in the inset.  (c) The second derivative of the map in (b) in the momentum direction.
}
\label{Fig1}
\end{figure*}

Figure \ref{Fig1}(a) shows an ARPES intensity map as a function of energy and momentum in the $\Gamma$--Y direction (see inset) for optimally doped Bi2212 at 20K.  Two main features are apparent at $\sim$50 and $\sim$140 meV (see red arrows), both characterized by a suppression of spectral weight, and the latter feature characterized by an inflection point in the dispersion.  The suppression of spectral weight becomes clearer at room temperature due to the broadening of the line shape [see Fig. \ref{Fig1}(b)].  Rather than being localized to a particular momentum or energy, the suppression extends along two lines cutting diagonally across the valence band [see the second derivative in the momentum direction in Fig. \ref{Fig1}(c)].  Each line causes a characteristic distortion where it crosses the valence band.

%%%%%%%%%%%%%%%%%%%%%%%%%%%%%%%%%
%Fig2
%%%%%%%%%%%%%%%%%%%%%%%%%%%%%%%%%
\begin{figure}\centering\includegraphics[width=3.375in]{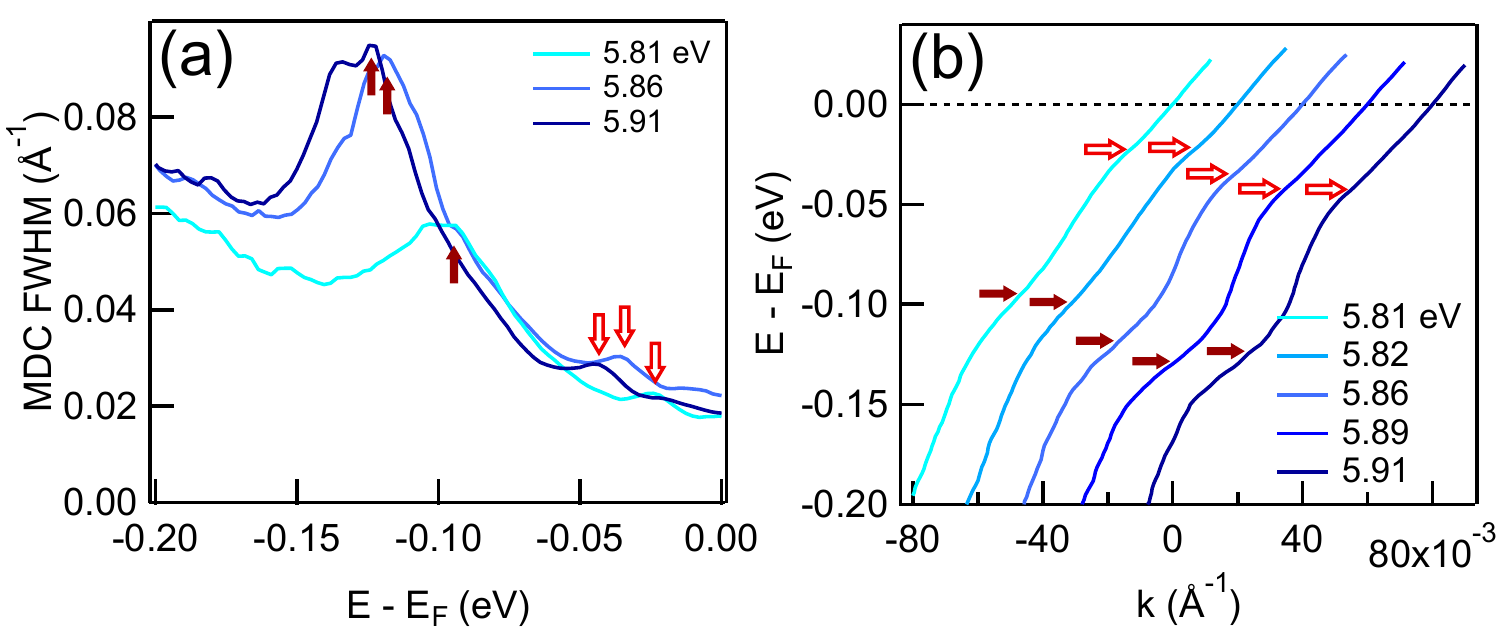}
\caption{(Color online)
(a) MDC full width at half maxima (FWHM) of optimally doped Bi2212 at 20K along $\Gamma$--Y, as measured by various photon energies.  (b) MDC dispersions, with measurements from different photon energies horizontally offset from each other.  The red arrows in both (a) and (b) indicate peaks in the MDC widths.
}
\label{Fig2}
\end{figure}

In Figure \ref{Fig2}, we report the nodal momentum distribution curve (MDC) dispersions and MDC widths at 20 K for various photon energies.  The MDCs are fitted using the standard Lorentzian function procedure\cite{LaShell2000}.  In the MDC widths shown in Fig. \ref{Fig2}(a), two clear peaks can be identified in the ranges of 20--50 and 100--140 meV, as expected from the lines of suppressed spectral weight observed in Fig. \ref{Fig1}.  At all photon energies, the peak in MDC widths at 100--140 meV roughly corresponds to an inflection point shown in Fig. \ref{Fig2}(b).  Although such an inflection point superficially resembles the renormalization of the electronic structure associated with electron-boson coupling\cite{LaShell2000}, such as the 70 meV kink in cuprates\cite{Lanzara2001}, its photon energy dependence points to a fundamentally different nature.  Furthermore, the renormalization from electron-boson coupling typically causes a step in the MDC widths, which is in contrast to the observed peaks in MDC widths.

%%%%%%%%%%%%%%%%%%%%%%%%%%%%%%%%%
%Fig3
%%%%%%%%%%%%%%%%%%%%%%%%%%%%%%%%%
\begin{figure}\centering\includegraphics[width=3.375in]{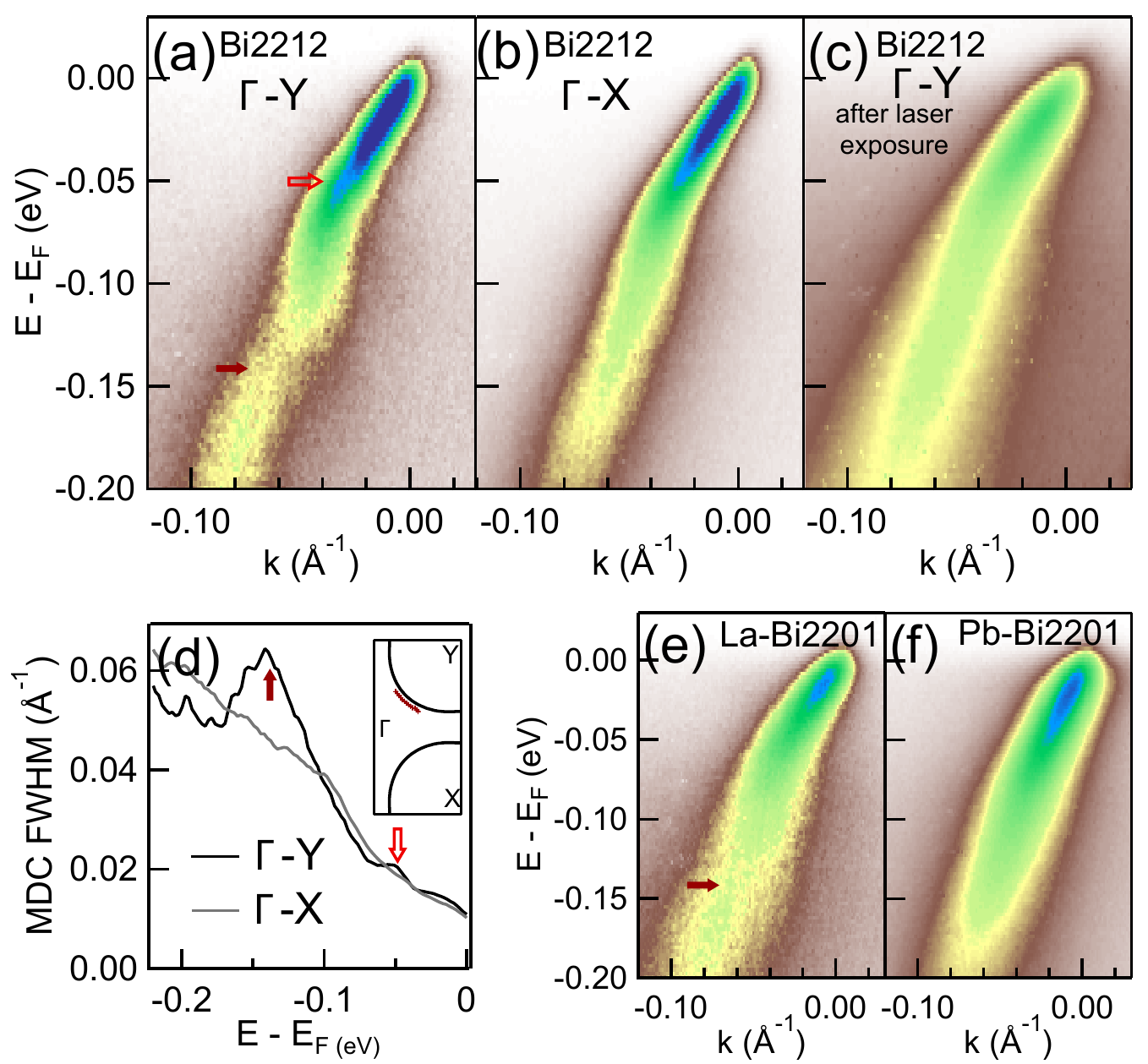}
\caption{(Color online) Raw ARPES intensity maps of Bi2212 at 20 K along the $\Gamma$--Y direction (a), the $\Gamma$--X direction (b), and again along $\Gamma$--Y after 4 h of laser exposure (c). (c) is broader than (a, b) due to sample differences rather than due to laser exposure.  (d) The MDC widths of the maps in (a, b).  The inset shows half of the first Brillouin zone, and the momentum locations of the observed suppression of spectral weight.  (e, f) Raw ARPES intensity maps of La-Bi2201 and overdoped Pb-Bi2201.
}
\label{Fig3}
\end{figure}

To better understand the origin of the suppression of spectral weight, in Fig. \ref{Fig3} we show data for different samples and experimental conditions.  In panels (a) and (b), we compare low-temperature measurements of the same double-layer Bi2212 sample along the $\Gamma$--Y and $\Gamma$--X directions respectively.  The distortion from suppression of spectral weight is clear along the $\Gamma$--Y direction but absent along the $\Gamma$--X direction.  Similarly, the suppression disappears in another sample along $\Gamma$--Y after 4 h of laser exposure on the same spot of the sample surface [see Figs. \ref{Fig3}(c)].  The intensity suppression does not disappear after a similar period of aging unless the sample was also exposed to light; the data shown in Fig. \ref{Fig2} were from a sample that had been aged for a few days but not exposed to light.  The disappearance of the suppression of spectral weight is even more stark when the MDC widths are compared along the $\Gamma$--Y and $\Gamma$--X directions [see Figs \ref{Fig3}(d)]. Peaks in the MDC widths appear along and near the $\Gamma$--Y direction at 50 and 140 meV, but not along the $\Gamma$--X direction.  The only feature along $\Gamma$--X is a broad step at 70 meV that is consistent with the electron-boson coupling widely studied in the literature\cite{Lanzara2001,Zhang2008}.

The primary distinction between the $\Gamma$--X and $\Gamma$--Y directions is the existence of an incommensurate superstructure modulation along $\Gamma$--Y\cite{Withers1988,He2008}, suggesting a relation between the suppression of spectral weight and the superstructure.  This might also explain the absence of the suppression in the laser-exposed sample, as the laser may destroy the superstructure near the surface.  Lastly, the importance of the superstructure is corroborated by looking at single-layered La-Bi2201 and Pb-Bi2201 samples along the (0,0)--($\pi$,$\pi$) direction [see Figs. \ref{Fig3}(e) and \ref{Fig3}(f)].  Doping with Pb is known to remove the superstructure\cite{Damascelli2003}; accordingly, a weak suppression of intensity is seen near 140 meV in La-Bi2201, and no similar feature is seen in Pb-Bi2201.  Note that since we observe the feature in single-layer La-Bi2201, which unlike Bi2212 has no bilayer band splitting\cite{Yamasaki2007}, this rules out any explanation based on intensity jumping between bilayer bands.

%%%%%%%%%%%%%%%%%%%%%%%%%%%%%%%%%
%Fig4
%%%%%%%%%%%%%%%%%%%%%%%%%%%%%%%%%
\begin{figure*}\centering\includegraphics[width=7in]{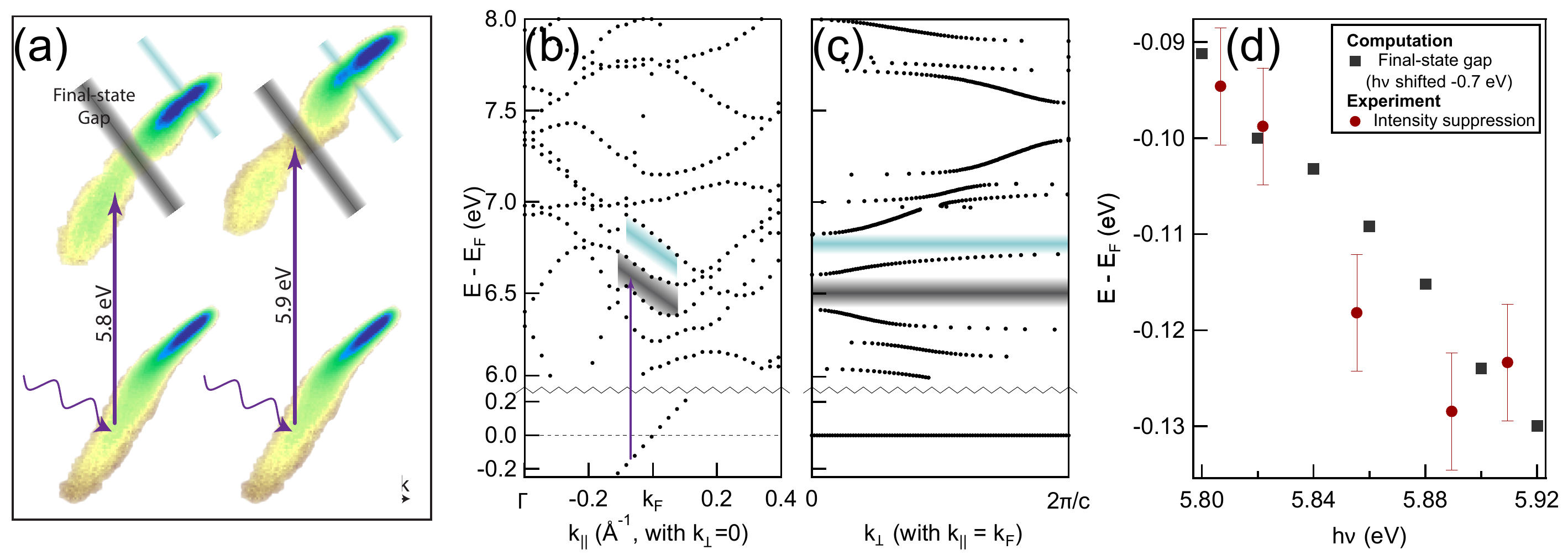}
\caption{(Color online) (a) An illustration of how the valence band absorbs a photon, and is distorted by the presence of two final-state gaps.  Two different photon energies are shown to illustrate how this affects the location of the distortion relative to the valence band.  (b) The dispersion of Bi2212 along the $\Gamma$--Y direction, with $k_\perp$ held at zero.  (c) The $k_\perp$ dispersion of Bi2212 at the node.  The gray bars in (b) and (c) indicate the main candidate for the lower final-state gap, while the lighter blue bars indicate a candidate for the upper final-state gap.  (d) A comparison of the calculated energy of the lower final-state gap to the energy where intensity suppression is found experimentally, by looking at peaks in the MDC widths.  The absolute computed final-state energies have been shifted by -0.7 eV to best fit the experimental data.  Error bars on the experimental measurements are estimated by the difference between the energy of greatest MDC width, and the inflection point of the MDC dispersion.}
\label{Fig4}
\end{figure*}

In Fig. \ref{Fig4}(a), we illustrate a possible mechanism for the dispersion anomaly here discussed.  As in a typical ARPES experiment, electrons in the material absorb photons, and escape from the sample, leaving behind holelike quasiparticle excitations.  But if we also account for the final states of the electrons after they have absorbed photons, some electrons with specific energy and momenta will fall in the space between unoccupied bands; this final-state gap falls along a line in momentum and energy.  Photoemission within the final-state gap is suppressed due to a lack of states, although not suppressed completely, due to the finite widths of the final states.  Finally, when the photon energy is varied, the final-state gap crosses the quasiparticle dispersion at a different point, in line with the observed photon energy dependence of the spectral weight suppression.

To corroborate this mechanism, we show fully relativistic one-step calculations of the electron final states of Bi2212 in Figs. \ref{Fig4}(b) and \ref{Fig4}(c).  We find that there is a final-state gap near the node, $\sim$6.5 eV above the Fermi energy, which qualitatively has the correct momentum dependence along the $\Gamma$--Y direction.  Furthermore, the electrons cannot get around the final-state gap through $k_\perp$, since the nodal quasiparticles have nearly no $k_\perp$ dispersion, and the gap exists at all values of $k_\perp$.   A similar final-state gap is also predicted near 6.75 eV, which explains the two features of intensity suppression in Bi2212.  Lastly, in Fig. \ref{Fig4}(d), we show that there is quantitative agreement between observations of the intensity suppression near 140 meV, and calculations of the lower final-state gap.  The absolute energies of the calculations are shifted by 0.7 eV, keeping in mind the inherent uncertainty in the absolute energies of first-principles band structure calculations.

%Discussion

Although the simulations are overall in good agreement with the experimental data, they do not account for the superstructure modulation, which appears to be relevant in the data.  The relevance of the superstructure is suggested by three observations: (a) the intensity suppression is present only along $\Gamma$--Y and not $\Gamma$--X; (b) the intensity suppression persists in single-layer La-Bi2212, but disappears upon Pb substitution (known to remove the superstructure); and (c) the intensity suppression disappears upon laser exposure.  So, while the experiments seem to suggest that the final-state gaps are tied to the presence of the superstructure modulation, the simulations clearly show that the final-state gaps exist even without the superstructure.  We note that in the calculations, the gap can only be observed if the imaginary part of the self-energy (i.e., half the width of the intensity peaks) is less than 0.4 eV.  Therefore, one possible explanation for the data is that the superstructure has no effect on the final-state gap in the $\Gamma$--Y direction, but broadens the gap in the $\Gamma$--X direction, rendering it invisible.  Within this picture, the overdoped Pb-Bi2201 and the laser-exposed Bi2212 may also have imaginary self-energy greater than 0.4 eV near the final-state gap, rendering it invisible in those samples as well.  In any case, the vanishing property of the dispersion anomaly may explain why it has not been shown in previous literature (although one study observed a similar feature and did not comment on it\cite{Perfetti2007}).  In an ordinary ARPES experiment, the anomaly will be quickly destroyed by the laser exposure, and thus the anomaly may at first appear unreproducible.  The disappearance of the anomaly may also have practical use, as some experiments may wish to focus on, for example, the 70 meV electron-boson kink, which is somewhat confounded when the dispersion anomaly is present.

Our observation of a final-state gap in Bi2212 and La-Bi2201 provides a model for addressing similar features in other materials with laser ARPES.  While final-state gaps have been observed with non-laser-based ARPES, observations with laser-ARPES open up new possibilities.  Indeed, while other light sources can probe unoccupied states high above the Fermi energy, laser ARPES probes energies between the sample's workfunction and 6--7 eV above the Fermi energy.  It is also possible that the improved bulk sensitivity of laser ARPES resolves the final-state gaps more sharply, since it would reduce the role of evanescent surface states\cite{Courths1989}.  Lastly, the greater momentum resolution of laser ARPES can be used to better resolve the dispersion of the final-state gap, or the final-state gap may be used to resolve the $k_\perp$ dispersion.

\begin{acknowledgments}
The angle-resolved photoemission spectroscopy part of this work was supported by Lawrence Berkeley National Laboratory's program on Quantum Materials, funded by the U.S. Department of Energy (DOE), Office of Science, Office of Basic Energy Sciences, Materials Sciences and Engineering Division, under Contract No. DE-AC02-05CH11231. The work at Northeastern University (NU) was supported by the DOE grant number DE-FG02-07ER46352, and benefited from NU's Advanced Scientific Computation Center (ASCC) and the NERSC supercomputing center through DOE grant number DE-AC02-05CH11231. The computational part of this work benefited from grid computing software provided by Techila Technologies Ltd.
\end{acknowledgments}

\bibliographystyle{apsrev4-1}
%\bibliographystyle{naturemag}
%\bibliography{tlmbib}
\bibliography{tlmbib}

\begin{thebibliography}{32}%
\makeatletter
\providecommand \@ifxundefined [1]{%
 \@ifx{#1\undefined}
}%
\providecommand \@ifnum [1]{%
 \ifnum #1\expandafter \@firstoftwo
 \else \expandafter \@secondoftwo
 \fi
}%
\providecommand \@ifx [1]{%
 \ifx #1\expandafter \@firstoftwo
 \else \expandafter \@secondoftwo
 \fi
}%
\providecommand \natexlab [1]{#1}%
\providecommand \enquote  [1]{``#1''}%
\providecommand \bibnamefont  [1]{#1}%
\providecommand \bibfnamefont [1]{#1}%
\providecommand \citenamefont [1]{#1}%
\providecommand \href@noop [0]{\@secondoftwo}%
\providecommand \href [0]{\begingroup \@sanitize@url \@href}%
\providecommand \@href[1]{\@@startlink{#1}\@@href}%
\providecommand \@@href[1]{\endgroup#1\@@endlink}%
\providecommand \@sanitize@url [0]{\catcode `\\12\catcode `\$12\catcode
  `\&12\catcode `\#12\catcode `\^12\catcode `\_12\catcode `\%12\relax}%
\providecommand \@@startlink[1]{}%
\providecommand \@@endlink[0]{}%
\providecommand \url  [0]{\begingroup\@sanitize@url \@url }%
\providecommand \@url [1]{\endgroup\@href {#1}{\urlprefix }}%
\providecommand \urlprefix  [0]{URL }%
\providecommand \Eprint [0]{\href }%
\providecommand \doibase [0]{http://dx.doi.org/}%
\providecommand \selectlanguage [0]{\@gobble}%
\providecommand \bibinfo  [0]{\@secondoftwo}%
\providecommand \bibfield  [0]{\@secondoftwo}%
\providecommand \translation [1]{[#1]}%
\providecommand \BibitemOpen [0]{}%
\providecommand \bibitemStop [0]{}%
\providecommand \bibitemNoStop [0]{.\EOS\space}%
\providecommand \EOS [0]{\spacefactor3000\relax}%
\providecommand \BibitemShut  [1]{\csname bibitem#1\endcsname}%
\let\auto@bib@innerbib\@empty
%</preamble>
\bibitem [{\citenamefont {H\"ufner}(2003)}]{Hufner2003}%
  \BibitemOpen
  \bibfield  {author} {\bibinfo {author} {\bibfnamefont {S.}~\bibnamefont
  {H\"ufner}},\ }\href@noop {} {\emph {\bibinfo {title} {Photoelectron
  Spectroscopy: Principles and Applications}}},\ \bibinfo {edition} {3rd}\ ed.\
  (\bibinfo  {publisher} {Springer},\ \bibinfo {year} {2003})\BibitemShut
  {NoStop}%
\bibitem [{\citenamefont {Damascelli}\ \emph {et~al.}(2003)\citenamefont
  {Damascelli}, \citenamefont {Hussain},\ and\ \citenamefont
  {Shen}}]{Damascelli2003}%
  \BibitemOpen
  \bibfield  {author} {\bibinfo {author} {\bibfnamefont {A.}~\bibnamefont
  {Damascelli}}, \bibinfo {author} {\bibfnamefont {Z.}~\bibnamefont {Hussain}},
  \ and\ \bibinfo {author} {\bibfnamefont {Z.-X.}\ \bibnamefont {Shen}},\
  }\href {\doibase 10.1103/RevModPhys.75.473} {\bibfield  {journal} {\bibinfo
  {journal} {Rev. Mod. Phys.}\ }\textbf {\bibinfo {volume} {75}},\ \bibinfo
  {pages} {473} (\bibinfo {year} {2003})}\BibitemShut {NoStop}%
\bibitem [{\citenamefont {Dietz}\ and\ \citenamefont
  {Eastman}(1978)}]{Dietz1978}%
  \BibitemOpen
  \bibfield  {author} {\bibinfo {author} {\bibfnamefont {E.}~\bibnamefont
  {Dietz}}\ and\ \bibinfo {author} {\bibfnamefont {D.~E.}\ \bibnamefont
  {Eastman}},\ }\href {\doibase 10.1103/PhysRevLett.41.1674} {\bibfield
  {journal} {\bibinfo  {journal} {Phys. Rev. Lett.}\ }\textbf {\bibinfo
  {volume} {41}},\ \bibinfo {pages} {1674} (\bibinfo {year}
  {1978})}\BibitemShut {NoStop}%
\bibitem [{\citenamefont {Courths}\ \emph {et~al.}(1989)\citenamefont
  {Courths}, \citenamefont {Wern}, \citenamefont {Leschik},\ and\ \citenamefont
  {H\"ufner}}]{Courths1989}%
  \BibitemOpen
  \bibfield  {author} {\bibinfo {author} {\bibfnamefont {R.}~\bibnamefont
  {Courths}}, \bibinfo {author} {\bibfnamefont {H.}~\bibnamefont {Wern}},
  \bibinfo {author} {\bibfnamefont {G.}~\bibnamefont {Leschik}}, \ and\
  \bibinfo {author} {\bibfnamefont {S.}~\bibnamefont {H\"ufner}},\ }\bibfield
  {booktitle} {\emph {\bibinfo {booktitle} {Zeitschrift für Physik B Condensed
  Matter}},\ }\href {\doibase 10.1007/BF01307390} {\ \textbf {\bibinfo {volume}
  {74}},\ \bibinfo {pages} {233} (\bibinfo {year} {1989})}\BibitemShut
  {NoStop}%
\bibitem [{\citenamefont {Strocov}\ \emph {et~al.}(1998)\citenamefont
  {Strocov}, \citenamefont {Claessen}, \citenamefont {Nicolay}, \citenamefont
  {H\"ufner}, \citenamefont {Kimura}, \citenamefont {Harasawa}, \citenamefont
  {Shin}, \citenamefont {Kakizaki}, \citenamefont {Nilsson}, \citenamefont
  {Starnberg},\ and\ \citenamefont {Blaha}}]{Strocov1998}%
  \BibitemOpen
  \bibfield  {author} {\bibinfo {author} {\bibfnamefont {V.~N.}\ \bibnamefont
  {Strocov}}, \bibinfo {author} {\bibfnamefont {R.}~\bibnamefont {Claessen}},
  \bibinfo {author} {\bibfnamefont {G.}~\bibnamefont {Nicolay}}, \bibinfo
  {author} {\bibfnamefont {S.}~\bibnamefont {H\"ufner}}, \bibinfo {author}
  {\bibfnamefont {A.}~\bibnamefont {Kimura}}, \bibinfo {author} {\bibfnamefont
  {A.}~\bibnamefont {Harasawa}}, \bibinfo {author} {\bibfnamefont
  {S.}~\bibnamefont {Shin}}, \bibinfo {author} {\bibfnamefont {A.}~\bibnamefont
  {Kakizaki}}, \bibinfo {author} {\bibfnamefont {P.~O.}\ \bibnamefont
  {Nilsson}}, \bibinfo {author} {\bibfnamefont {H.~I.}\ \bibnamefont
  {Starnberg}}, \ and\ \bibinfo {author} {\bibfnamefont {P.}~\bibnamefont
  {Blaha}},\ }\href {\doibase 10.1103/PhysRevLett.81.4943} {\bibfield
  {journal} {\bibinfo  {journal} {Phys. Rev. Lett.}\ }\textbf {\bibinfo
  {volume} {81}},\ \bibinfo {pages} {4943} (\bibinfo {year}
  {1998})}\BibitemShut {NoStop}%
\bibitem [{\citenamefont {Strocov}\ \emph {et~al.}(2012)\citenamefont
  {Strocov}, \citenamefont {Shi}, \citenamefont {Kobayashi}, \citenamefont
  {Monney}, \citenamefont {Wang}, \citenamefont {Krempasky}, \citenamefont
  {Schmitt}, \citenamefont {Patthey}, \citenamefont {Berger},\ and\
  \citenamefont {Blaha}}]{Strocov2012}%
  \BibitemOpen
  \bibfield  {author} {\bibinfo {author} {\bibfnamefont {V.~N.}\ \bibnamefont
  {Strocov}}, \bibinfo {author} {\bibfnamefont {M.}~\bibnamefont {Shi}},
  \bibinfo {author} {\bibfnamefont {M.}~\bibnamefont {Kobayashi}}, \bibinfo
  {author} {\bibfnamefont {C.}~\bibnamefont {Monney}}, \bibinfo {author}
  {\bibfnamefont {X.}~\bibnamefont {Wang}}, \bibinfo {author} {\bibfnamefont
  {J.}~\bibnamefont {Krempasky}}, \bibinfo {author} {\bibfnamefont
  {T.}~\bibnamefont {Schmitt}}, \bibinfo {author} {\bibfnamefont
  {L.}~\bibnamefont {Patthey}}, \bibinfo {author} {\bibfnamefont
  {H.}~\bibnamefont {Berger}}, \ and\ \bibinfo {author} {\bibfnamefont
  {P.}~\bibnamefont {Blaha}},\ }\href {\doibase 10.1103/PhysRevLett.109.086401}
  {\bibfield  {journal} {\bibinfo  {journal} {Phys. Rev. Lett.}\ }\textbf
  {\bibinfo {volume} {109}},\ \bibinfo {pages} {086401} (\bibinfo {year}
  {2012})}\BibitemShut {NoStop}%
\bibitem [{\citenamefont {Pendry}(1976)}]{Pendry1976}%
  \BibitemOpen
  \bibfield  {author} {\bibinfo {author} {\bibfnamefont {J.}~\bibnamefont
  {Pendry}},\ }\href {\doibase 10.1016/0039-6028(76)90355-1} {\bibfield
  {journal} {\bibinfo  {journal} {Surface Science}\ }\textbf {\bibinfo {volume}
  {57}},\ \bibinfo {pages} {679} (\bibinfo {year} {1976})}\BibitemShut
  {NoStop}%
\bibitem [{\citenamefont {Yoshida}(2013)}]{Yoshida2013}%
  \BibitemOpen
  \bibfield  {author} {\bibinfo {author} {\bibfnamefont {H.}~\bibnamefont
  {Yoshida}},\ }\href {\doibase 10.1063/1.4822119} {\bibfield  {journal}
  {\bibinfo  {journal} {Rev. Sci. Instrum.}\ }\textbf {\bibinfo {volume}
  {84}},\ \bibinfo {pages} {103901} (\bibinfo {year} {2013})}\BibitemShut
  {NoStop}%
\bibitem [{\citenamefont {Lindroos}\ \emph {et~al.}(1986)\citenamefont
  {Lindroos}, \citenamefont {Pfn\"ur},\ and\ \citenamefont
  {Menzel}}]{Lindroos1986}%
  \BibitemOpen
  \bibfield  {author} {\bibinfo {author} {\bibfnamefont {M.}~\bibnamefont
  {Lindroos}}, \bibinfo {author} {\bibfnamefont {H.}~\bibnamefont {Pfn\"ur}}, \
  and\ \bibinfo {author} {\bibfnamefont {D.}~\bibnamefont {Menzel}},\ }\href
  {\doibase 10.1103/PhysRevB.33.6684} {\bibfield  {journal} {\bibinfo
  {journal} {Phys. Rev. B}\ }\textbf {\bibinfo {volume} {33}},\ \bibinfo
  {pages} {6684} (\bibinfo {year} {1986})}\BibitemShut {NoStop}%
\bibitem [{\citenamefont {Lindroos}\ \emph
  {et~al.}(1987{\natexlab{a}})\citenamefont {Lindroos}, \citenamefont
  {Pfn\"ur},\ and\ \citenamefont {Menzel}}]{Lindroos1987}%
  \BibitemOpen
  \bibfield  {author} {\bibinfo {author} {\bibfnamefont {M.}~\bibnamefont
  {Lindroos}}, \bibinfo {author} {\bibfnamefont {H.}~\bibnamefont {Pfn\"ur}}, \
  and\ \bibinfo {author} {\bibfnamefont {D.}~\bibnamefont {Menzel}},\ }\href
  {\doibase 10.1016/S0039-6028(87)81137-8} {\bibfield  {journal} {\bibinfo
  {journal} {Surface Science}\ }\textbf {\bibinfo {volume} {192}},\ \bibinfo
  {pages} {421} (\bibinfo {year} {1987}{\natexlab{a}})}\BibitemShut {NoStop}%
\bibitem [{\citenamefont {Lindroos}\ \emph
  {et~al.}(1987{\natexlab{b}})\citenamefont {Lindroos}, \citenamefont
  {Pfn\"ur}, \citenamefont {Feulner},\ and\ \citenamefont
  {Menzel}}]{Lindroos1987a}%
  \BibitemOpen
  \bibfield  {author} {\bibinfo {author} {\bibfnamefont {M.}~\bibnamefont
  {Lindroos}}, \bibinfo {author} {\bibfnamefont {H.}~\bibnamefont {Pfn\"ur}},
  \bibinfo {author} {\bibfnamefont {P.}~\bibnamefont {Feulner}}, \ and\
  \bibinfo {author} {\bibfnamefont {D.}~\bibnamefont {Menzel}},\ }\href
  {\doibase 10.1016/0039-6028(87)90046-X} {\bibfield  {journal} {\bibinfo
  {journal} {Surface Science}\ }\textbf {\bibinfo {volume} {180}},\ \bibinfo
  {pages} {237} (\bibinfo {year} {1987}{\natexlab{b}})}\BibitemShut {NoStop}%
\bibitem [{\citenamefont {Strocov}\ \emph {et~al.}(2003)\citenamefont
  {Strocov}, \citenamefont {Claessen},\ and\ \citenamefont
  {Blaha}}]{Strocov2003}%
  \BibitemOpen
  \bibfield  {author} {\bibinfo {author} {\bibfnamefont {V.~N.}\ \bibnamefont
  {Strocov}}, \bibinfo {author} {\bibfnamefont {R.}~\bibnamefont {Claessen}}, \
  and\ \bibinfo {author} {\bibfnamefont {P.}~\bibnamefont {Blaha}},\ }\href
  {\doibase 10.1103/PhysRevB.68.144509} {\bibfield  {journal} {\bibinfo
  {journal} {Phys. Rev. B}\ }\textbf {\bibinfo {volume} {68}},\ \bibinfo
  {pages} {144509} (\bibinfo {year} {2003})}\BibitemShut {NoStop}%
\bibitem [{\citenamefont {Koralek}\ \emph {et~al.}(2007)\citenamefont
  {Koralek}, \citenamefont {Douglas}, \citenamefont {Plumb}, \citenamefont
  {Griffith}, \citenamefont {Cundiff}, \citenamefont {Kapteyn}, \citenamefont
  {Murnane},\ and\ \citenamefont {Dessau}}]{Koralek2007}%
  \BibitemOpen
  \bibfield  {author} {\bibinfo {author} {\bibfnamefont {J.~D.}\ \bibnamefont
  {Koralek}}, \bibinfo {author} {\bibfnamefont {J.~F.}\ \bibnamefont
  {Douglas}}, \bibinfo {author} {\bibfnamefont {N.~C.}\ \bibnamefont {Plumb}},
  \bibinfo {author} {\bibfnamefont {J.~D.}\ \bibnamefont {Griffith}}, \bibinfo
  {author} {\bibfnamefont {S.~T.}\ \bibnamefont {Cundiff}}, \bibinfo {author}
  {\bibfnamefont {H.~C.}\ \bibnamefont {Kapteyn}}, \bibinfo {author}
  {\bibfnamefont {M.~M.}\ \bibnamefont {Murnane}}, \ and\ \bibinfo {author}
  {\bibfnamefont {D.~S.}\ \bibnamefont {Dessau}},\ }\href {\doibase
  10.1063/1.2722413} {\bibfield  {journal} {\bibinfo  {journal} {Rev. Sci.
  Instrum.}\ }\textbf {\bibinfo {volume} {78}},\ \bibinfo {pages} {053905}
  (\bibinfo {year} {2007})}\BibitemShut {NoStop}%
\bibitem [{\citenamefont {Liu}\ \emph {et~al.}(2008)\citenamefont {Liu},
  \citenamefont {Wang}, \citenamefont {Zhu}, \citenamefont {Zhang},
  \citenamefont {Zhang}, \citenamefont {Wang}, \citenamefont {Zhou},
  \citenamefont {Zhang}, \citenamefont {Liu}, \citenamefont {Zhao},
  \citenamefont {Meng}, \citenamefont {Dong}, \citenamefont {Chen},
  \citenamefont {Xu},\ and\ \citenamefont {Zhou}}]{Liu2008}%
  \BibitemOpen
  \bibfield  {author} {\bibinfo {author} {\bibfnamefont {G.}~\bibnamefont
  {Liu}}, \bibinfo {author} {\bibfnamefont {G.}~\bibnamefont {Wang}}, \bibinfo
  {author} {\bibfnamefont {Y.}~\bibnamefont {Zhu}}, \bibinfo {author}
  {\bibfnamefont {H.}~\bibnamefont {Zhang}}, \bibinfo {author} {\bibfnamefont
  {G.}~\bibnamefont {Zhang}}, \bibinfo {author} {\bibfnamefont
  {X.}~\bibnamefont {Wang}}, \bibinfo {author} {\bibfnamefont {Y.}~\bibnamefont
  {Zhou}}, \bibinfo {author} {\bibfnamefont {W.}~\bibnamefont {Zhang}},
  \bibinfo {author} {\bibfnamefont {H.}~\bibnamefont {Liu}}, \bibinfo {author}
  {\bibfnamefont {L.}~\bibnamefont {Zhao}}, \bibinfo {author} {\bibfnamefont
  {J.}~\bibnamefont {Meng}}, \bibinfo {author} {\bibfnamefont {X.}~\bibnamefont
  {Dong}}, \bibinfo {author} {\bibfnamefont {C.}~\bibnamefont {Chen}}, \bibinfo
  {author} {\bibfnamefont {Z.}~\bibnamefont {Xu}}, \ and\ \bibinfo {author}
  {\bibfnamefont {X.~J.}\ \bibnamefont {Zhou}},\ }\href {\doibase
  10.1063/1.2835901} {\bibfield  {journal} {\bibinfo  {journal} {Rev. Sci.
  Instrum.}\ }\textbf {\bibinfo {volume} {79}},\ \bibinfo {pages} {023105}
  (\bibinfo {year} {2008})}\BibitemShut {NoStop}%
\bibitem [{\citenamefont {Kiss}\ \emph {et~al.}(2008)\citenamefont {Kiss},
  \citenamefont {Shimojima}, \citenamefont {Ishizaka}, \citenamefont
  {Chainani}, \citenamefont {Togashi}, \citenamefont {Kanai}, \citenamefont
  {Wang}, \citenamefont {Chen}, \citenamefont {Watanabe},\ and\ \citenamefont
  {Shin}}]{Kiss2008}%
  \BibitemOpen
  \bibfield  {author} {\bibinfo {author} {\bibfnamefont {T.}~\bibnamefont
  {Kiss}}, \bibinfo {author} {\bibfnamefont {T.}~\bibnamefont {Shimojima}},
  \bibinfo {author} {\bibfnamefont {K.}~\bibnamefont {Ishizaka}}, \bibinfo
  {author} {\bibfnamefont {A.}~\bibnamefont {Chainani}}, \bibinfo {author}
  {\bibfnamefont {T.}~\bibnamefont {Togashi}}, \bibinfo {author} {\bibfnamefont
  {T.}~\bibnamefont {Kanai}}, \bibinfo {author} {\bibfnamefont {X.-Y.}\
  \bibnamefont {Wang}}, \bibinfo {author} {\bibfnamefont {C.-T.}\ \bibnamefont
  {Chen}}, \bibinfo {author} {\bibfnamefont {S.}~\bibnamefont {Watanabe}}, \
  and\ \bibinfo {author} {\bibfnamefont {S.}~\bibnamefont {Shin}},\ }\href
  {\doibase 10.1063/1.2839010} {\bibfield  {journal} {\bibinfo  {journal} {Rev.
  Sci. Instrum.}\ }\textbf {\bibinfo {volume} {79}},\ \bibinfo {pages} {023106}
  (\bibinfo {year} {2008})}\BibitemShut {NoStop}%
\bibitem [{\citenamefont {Smallwood}\ \emph {et~al.}(2012)\citenamefont
  {Smallwood}, \citenamefont {Jozwiak}, \citenamefont {Zhang},\ and\
  \citenamefont {Lanzara}}]{Smallwood2012RSI}%
  \BibitemOpen
  \bibfield  {author} {\bibinfo {author} {\bibfnamefont {C.~L.}\ \bibnamefont
  {Smallwood}}, \bibinfo {author} {\bibfnamefont {C.}~\bibnamefont {Jozwiak}},
  \bibinfo {author} {\bibfnamefont {W.}~\bibnamefont {Zhang}}, \ and\ \bibinfo
  {author} {\bibfnamefont {A.}~\bibnamefont {Lanzara}},\ }\href {\doibase
  10.1063/1.4772070} {\bibfield  {journal} {\bibinfo  {journal} {Rev. Sci.
  Instrum.}\ }\textbf {\bibinfo {volume} {83}},\ \bibinfo {pages} {123904}
  (\bibinfo {year} {2012})}\BibitemShut {NoStop}%
\bibitem [{\citenamefont {\"Arr\"al\"a}\ \emph {et~al.}(2013)\citenamefont
  {\"Arr\"al\"a}, \citenamefont {Nieminen}, \citenamefont {Braun},
  \citenamefont {Ebert},\ and\ \citenamefont {Lindroos}}]{Arrala2013}%
  \BibitemOpen
  \bibfield  {author} {\bibinfo {author} {\bibfnamefont {M.}~\bibnamefont
  {\"Arr\"al\"a}}, \bibinfo {author} {\bibfnamefont {J.}~\bibnamefont
  {Nieminen}}, \bibinfo {author} {\bibfnamefont {J.}~\bibnamefont {Braun}},
  \bibinfo {author} {\bibfnamefont {H.}~\bibnamefont {Ebert}}, \ and\ \bibinfo
  {author} {\bibfnamefont {M.}~\bibnamefont {Lindroos}},\ }\href {\doibase
  10.1103/PhysRevB.88.195413} {\bibfield  {journal} {\bibinfo  {journal} {Phys.
  Rev. B}\ }\textbf {\bibinfo {volume} {88}},\ \bibinfo {pages} {195413}
  (\bibinfo {year} {2013})}\BibitemShut {NoStop}%
\bibitem [{\citenamefont {Rienks}\ \emph {et~al.}(2014)\citenamefont {Rienks},
  \citenamefont {\"Arr\"al\"a}, \citenamefont {Lindroos}, \citenamefont {Roth},
  \citenamefont {Tabis}, \citenamefont {Yu}, \citenamefont {Greven},\ and\
  \citenamefont {Fink}}]{Rienks2014}%
  \BibitemOpen
  \bibfield  {author} {\bibinfo {author} {\bibfnamefont {E.}~\bibnamefont
  {Rienks}}, \bibinfo {author} {\bibfnamefont {M.}~\bibnamefont
  {\"Arr\"al\"a}}, \bibinfo {author} {\bibfnamefont {M.}~\bibnamefont
  {Lindroos}}, \bibinfo {author} {\bibfnamefont {F.}~\bibnamefont {Roth}},
  \bibinfo {author} {\bibfnamefont {W.}~\bibnamefont {Tabis}}, \bibinfo
  {author} {\bibfnamefont {G.}~\bibnamefont {Yu}}, \bibinfo {author}
  {\bibfnamefont {M.}~\bibnamefont {Greven}}, \ and\ \bibinfo {author}
  {\bibfnamefont {J.}~\bibnamefont {Fink}},\ }\href {\doibase
  10.1103/PhysRevLett.113.137001} {\bibfield  {journal} {\bibinfo  {journal}
  {Phys. Rev. Lett.}\ }\textbf {\bibinfo {volume} {113}},\ \bibinfo {pages}
  {137001} (\bibinfo {year} {2014})}\BibitemShut {NoStop}%
\bibitem [{\citenamefont {Braun}(1996)}]{Braun1996}%
  \BibitemOpen
  \bibfield  {author} {\bibinfo {author} {\bibfnamefont {J.}~\bibnamefont
  {Braun}},\ }\href {\doibase 10.1088/0034-4885/59/10/002} {\bibfield
  {journal} {\bibinfo  {journal} {Rep. Prog. Phys.}\ }\textbf {\bibinfo
  {volume} {59}},\ \bibinfo {pages} {1267} (\bibinfo {year}
  {1996})}\BibitemShut {NoStop}%
\bibitem [{\citenamefont {Bansil}\ and\ \citenamefont
  {Lindroos}(1999)}]{Bansil1999}%
  \BibitemOpen
  \bibfield  {author} {\bibinfo {author} {\bibfnamefont {A.}~\bibnamefont
  {Bansil}}\ and\ \bibinfo {author} {\bibfnamefont {M.}~\bibnamefont
  {Lindroos}},\ }\href {\doibase 10.1103/PhysRevLett.83.5154} {\bibfield
  {journal} {\bibinfo  {journal} {Phys. Rev. Lett.}\ }\textbf {\bibinfo
  {volume} {83}},\ \bibinfo {pages} {5154} (\bibinfo {year}
  {1999})}\BibitemShut {NoStop}%
\bibitem [{\citenamefont {Lindroos}\ \emph {et~al.}(2002)\citenamefont
  {Lindroos}, \citenamefont {Sahrakorpi},\ and\ \citenamefont
  {Bansil}}]{Lindroos2002}%
  \BibitemOpen
  \bibfield  {author} {\bibinfo {author} {\bibfnamefont {M.}~\bibnamefont
  {Lindroos}}, \bibinfo {author} {\bibfnamefont {S.}~\bibnamefont
  {Sahrakorpi}}, \ and\ \bibinfo {author} {\bibfnamefont {A.}~\bibnamefont
  {Bansil}},\ }\href {\doibase 10.1103/PhysRevB.65.054514} {\bibfield
  {journal} {\bibinfo  {journal} {Phys. Rev. B}\ }\textbf {\bibinfo {volume}
  {65}},\ \bibinfo {pages} {054514} (\bibinfo {year} {2002})}\BibitemShut
  {NoStop}%
\bibitem [{\citenamefont {Sahrakorpi}\ \emph {et~al.}(2005)\citenamefont
  {Sahrakorpi}, \citenamefont {Lindroos}, \citenamefont {Markiewicz},\ and\
  \citenamefont {Bansil}}]{Sahrakorpi2005}%
  \BibitemOpen
  \bibfield  {author} {\bibinfo {author} {\bibfnamefont {S.}~\bibnamefont
  {Sahrakorpi}}, \bibinfo {author} {\bibfnamefont {M.}~\bibnamefont
  {Lindroos}}, \bibinfo {author} {\bibfnamefont {R.~S.}\ \bibnamefont
  {Markiewicz}}, \ and\ \bibinfo {author} {\bibfnamefont {A.}~\bibnamefont
  {Bansil}},\ }\href {\doibase 10.1103/PhysRevLett.95.157601} {\bibfield
  {journal} {\bibinfo  {journal} {Phys. Rev. Lett.}\ }\textbf {\bibinfo
  {volume} {95}},\ \bibinfo {pages} {157601} (\bibinfo {year}
  {2005})}\BibitemShut {NoStop}%
\bibitem [{\citenamefont {Malmstr{\"o}m}\ and\ \citenamefont
  {Rundgren}(1980)}]{Malmstrom1980}%
  \BibitemOpen
  \bibfield  {author} {\bibinfo {author} {\bibfnamefont {G.}~\bibnamefont
  {Malmstr{\"o}m}}\ and\ \bibinfo {author} {\bibfnamefont {J.}~\bibnamefont
  {Rundgren}},\ }\href {\doibase 10.1016/0010-4655(80)90053-3} {\bibfield
  {journal} {\bibinfo  {journal} {Comput. Phys. Commun.}\ }\textbf {\bibinfo
  {volume} {19}},\ \bibinfo {pages} {263 } (\bibinfo {year}
  {1980})}\BibitemShut {NoStop}%
\bibitem [{\citenamefont {Nieminen}\ \emph {et~al.}(2012)\citenamefont
  {Nieminen}, \citenamefont {Suominen}, \citenamefont {Das}, \citenamefont
  {Markiewicz},\ and\ \citenamefont {Bansil}}]{Nieminen2012}%
  \BibitemOpen
  \bibfield  {author} {\bibinfo {author} {\bibfnamefont {J.}~\bibnamefont
  {Nieminen}}, \bibinfo {author} {\bibfnamefont {I.}~\bibnamefont {Suominen}},
  \bibinfo {author} {\bibfnamefont {T.}~\bibnamefont {Das}}, \bibinfo {author}
  {\bibfnamefont {R.~S.}\ \bibnamefont {Markiewicz}}, \ and\ \bibinfo {author}
  {\bibfnamefont {A.}~\bibnamefont {Bansil}},\ }\href {\doibase
  10.1103/PhysRevB.85.214504} {\bibfield  {journal} {\bibinfo  {journal} {Phys.
  Rev. B}\ }\textbf {\bibinfo {volume} {85}},\ \bibinfo {pages} {214504}
  (\bibinfo {year} {2012})}\BibitemShut {NoStop}%
\bibitem [{\citenamefont {Blaha}\ \emph {et~al.}(2001)\citenamefont {Blaha},
  \citenamefont {Schwarz}, \citenamefont {Madsen}, \citenamefont {Kvasnicka},\
  and\ \citenamefont {Luitz}}]{Blaha2001}%
  \BibitemOpen
  \bibfield  {author} {\bibinfo {author} {\bibfnamefont {P.}~\bibnamefont
  {Blaha}}, \bibinfo {author} {\bibfnamefont {K.}~\bibnamefont {Schwarz}},
  \bibinfo {author} {\bibfnamefont {G.~K.~H.}\ \bibnamefont {Madsen}}, \bibinfo
  {author} {\bibfnamefont {D.}~\bibnamefont {Kvasnicka}}, \ and\ \bibinfo
  {author} {\bibfnamefont {J.}~\bibnamefont {Luitz}},\ }\href@noop {}
  {\bibfield  {journal} {\bibinfo  {journal} {WIEN2k, an Augmented Plane Wave
  Plus Local Orbitals Program for Calculating Crystal Properties (Karlheinz
  Schwarz, Techn. Universit\"at Wien, Austria), ISBN 3-9501031-1-2}\ }
  (\bibinfo {year} {2001})}\BibitemShut {NoStop}%
\bibitem [{\citenamefont {LaShell}\ \emph {et~al.}(2000)\citenamefont
  {LaShell}, \citenamefont {Jensen},\ and\ \citenamefont
  {Balasubramanian}}]{LaShell2000}%
  \BibitemOpen
  \bibfield  {author} {\bibinfo {author} {\bibfnamefont {S.}~\bibnamefont
  {LaShell}}, \bibinfo {author} {\bibfnamefont {E.}~\bibnamefont {Jensen}}, \
  and\ \bibinfo {author} {\bibfnamefont {T.}~\bibnamefont {Balasubramanian}},\
  }\href {\doibase 10.1103/PhysRevB.61.2371} {\bibfield  {journal} {\bibinfo
  {journal} {Phys. Rev. B}\ }\textbf {\bibinfo {volume} {61}},\ \bibinfo
  {pages} {2371} (\bibinfo {year} {2000})}\BibitemShut {NoStop}%
\bibitem [{\citenamefont {Lanzara}\ \emph {et~al.}(2001)\citenamefont
  {Lanzara}, \citenamefont {Bogdanov}, \citenamefont {Zhou}, \citenamefont
  {Kellar}, \citenamefont {Feng}, \citenamefont {Lu}, \citenamefont {Yoshida},
  \citenamefont {Eisaki}, \citenamefont {Fujimori}, \citenamefont {Kishio},
  \citenamefont {Shimoyama}, \citenamefont {Noda}, \citenamefont {Uchida},
  \citenamefont {Hussain},\ and\ \citenamefont {Shen}}]{Lanzara2001}%
  \BibitemOpen
  \bibfield  {author} {\bibinfo {author} {\bibfnamefont {A.}~\bibnamefont
  {Lanzara}}, \bibinfo {author} {\bibfnamefont {P.~V.}\ \bibnamefont
  {Bogdanov}}, \bibinfo {author} {\bibfnamefont {X.~J.}\ \bibnamefont {Zhou}},
  \bibinfo {author} {\bibfnamefont {S.~A.}\ \bibnamefont {Kellar}}, \bibinfo
  {author} {\bibfnamefont {D.~L.}\ \bibnamefont {Feng}}, \bibinfo {author}
  {\bibfnamefont {E.~D.}\ \bibnamefont {Lu}}, \bibinfo {author} {\bibfnamefont
  {T.}~\bibnamefont {Yoshida}}, \bibinfo {author} {\bibfnamefont
  {H.}~\bibnamefont {Eisaki}}, \bibinfo {author} {\bibfnamefont
  {A.}~\bibnamefont {Fujimori}}, \bibinfo {author} {\bibfnamefont
  {K.}~\bibnamefont {Kishio}}, \bibinfo {author} {\bibfnamefont {J.-I.}\
  \bibnamefont {Shimoyama}}, \bibinfo {author} {\bibfnamefont {T.}~\bibnamefont
  {Noda}}, \bibinfo {author} {\bibfnamefont {S.}~\bibnamefont {Uchida}},
  \bibinfo {author} {\bibfnamefont {Z.}~\bibnamefont {Hussain}}, \ and\
  \bibinfo {author} {\bibfnamefont {Z.-X.}\ \bibnamefont {Shen}},\ }\href
  {\doibase 10.1038/35087518} {\bibfield  {journal} {\bibinfo  {journal}
  {Nature}\ }\textbf {\bibinfo {volume} {412}},\ \bibinfo {pages} {510}
  (\bibinfo {year} {2001})}\BibitemShut {NoStop}%
\bibitem [{\citenamefont {Zhang}\ \emph {et~al.}(2008)\citenamefont {Zhang},
  \citenamefont {Liu}, \citenamefont {Zhao}, \citenamefont {Liu}, \citenamefont
  {Meng}, \citenamefont {Dong}, \citenamefont {Lu}, \citenamefont {Wen},
  \citenamefont {Xu}, \citenamefont {Gu}, \citenamefont {Sasagawa},
  \citenamefont {Wang}, \citenamefont {Zhu}, \citenamefont {Zhang},
  \citenamefont {Zhou}, \citenamefont {Wang}, \citenamefont {Zhao},
  \citenamefont {Chen}, \citenamefont {Xu},\ and\ \citenamefont
  {Zhou}}]{Zhang2008}%
  \BibitemOpen
  \bibfield  {author} {\bibinfo {author} {\bibfnamefont {W.~T.}\ \bibnamefont
  {Zhang}}, \bibinfo {author} {\bibfnamefont {G.~D.}\ \bibnamefont {Liu}},
  \bibinfo {author} {\bibfnamefont {L.}~\bibnamefont {Zhao}}, \bibinfo {author}
  {\bibfnamefont {H.~Y.}\ \bibnamefont {Liu}}, \bibinfo {author} {\bibfnamefont
  {J.~Q.}\ \bibnamefont {Meng}}, \bibinfo {author} {\bibfnamefont {X.~L.}\
  \bibnamefont {Dong}}, \bibinfo {author} {\bibfnamefont {W.}~\bibnamefont
  {Lu}}, \bibinfo {author} {\bibfnamefont {J.~S.}\ \bibnamefont {Wen}},
  \bibinfo {author} {\bibfnamefont {Z.~J.}\ \bibnamefont {Xu}}, \bibinfo
  {author} {\bibfnamefont {G.~D.}\ \bibnamefont {Gu}}, \bibinfo {author}
  {\bibfnamefont {T.}~\bibnamefont {Sasagawa}}, \bibinfo {author}
  {\bibfnamefont {G.~L.}\ \bibnamefont {Wang}}, \bibinfo {author}
  {\bibfnamefont {Y.}~\bibnamefont {Zhu}}, \bibinfo {author} {\bibfnamefont
  {H.~B.}\ \bibnamefont {Zhang}}, \bibinfo {author} {\bibfnamefont
  {Y.}~\bibnamefont {Zhou}}, \bibinfo {author} {\bibfnamefont {X.~Y.}\
  \bibnamefont {Wang}}, \bibinfo {author} {\bibfnamefont {Z.~X.}\ \bibnamefont
  {Zhao}}, \bibinfo {author} {\bibfnamefont {C.~T.}\ \bibnamefont {Chen}},
  \bibinfo {author} {\bibfnamefont {Z.~Y.}\ \bibnamefont {Xu}}, \ and\ \bibinfo
  {author} {\bibfnamefont {X.~J.}\ \bibnamefont {Zhou}},\ }\href {\doibase
  10.1103/PhysRevLett.100.107002} {\bibfield  {journal} {\bibinfo  {journal}
  {Phys. Rev. Lett.}\ }\textbf {\bibinfo {volume} {100}},\ \bibinfo {pages}
  {107002} (\bibinfo {year} {2008})}\BibitemShut {NoStop}%
\bibitem [{\citenamefont {Withers}\ \emph {et~al.}(1988)\citenamefont
  {Withers}, \citenamefont {Thompson}, \citenamefont {Wallenberg},
  \citenamefont {FitzGerald}, \citenamefont {Anderson},\ and\ \citenamefont
  {Hyde}}]{Withers1988}%
  \BibitemOpen
  \bibfield  {author} {\bibinfo {author} {\bibfnamefont {R.~L.}\ \bibnamefont
  {Withers}}, \bibinfo {author} {\bibfnamefont {J.~G.}\ \bibnamefont
  {Thompson}}, \bibinfo {author} {\bibfnamefont {L.~R.}\ \bibnamefont
  {Wallenberg}}, \bibinfo {author} {\bibfnamefont {J.~D.}\ \bibnamefont
  {FitzGerald}}, \bibinfo {author} {\bibfnamefont {J.~S.}\ \bibnamefont
  {Anderson}}, \ and\ \bibinfo {author} {\bibfnamefont {B.~G.}\ \bibnamefont
  {Hyde}},\ }\href {\doibase 10.1088/0022-3719/21/36/007} {\bibfield  {journal}
  {\bibinfo  {journal} {J. Phys. C}\ }\textbf {\bibinfo {volume} {21}},\
  \bibinfo {pages} {6067} (\bibinfo {year} {1988})}\BibitemShut {NoStop}%
\bibitem [{\citenamefont {He}\ \emph {et~al.}(2008)\citenamefont {He},
  \citenamefont {Graser}, \citenamefont {Hirschfeld},\ and\ \citenamefont
  {Cheng}}]{He2008}%
  \BibitemOpen
  \bibfield  {author} {\bibinfo {author} {\bibfnamefont {Y.}~\bibnamefont
  {He}}, \bibinfo {author} {\bibfnamefont {S.}~\bibnamefont {Graser}}, \bibinfo
  {author} {\bibfnamefont {P.~J.}\ \bibnamefont {Hirschfeld}}, \ and\ \bibinfo
  {author} {\bibfnamefont {H.-P.}\ \bibnamefont {Cheng}},\ }\href {\doibase
  10.1103/PhysRevB.77.220507} {\bibfield  {journal} {\bibinfo  {journal} {Phys.
  Rev. B}\ }\textbf {\bibinfo {volume} {77}},\ \bibinfo {pages} {220507}
  (\bibinfo {year} {2008})}\BibitemShut {NoStop}%
\bibitem [{\citenamefont {Yamasaki}\ \emph {et~al.}(2007)\citenamefont
  {Yamasaki}, \citenamefont {Yamazaki}, \citenamefont {Ino}, \citenamefont
  {Arita}, \citenamefont {Namatame}, \citenamefont {Taniguchi}, \citenamefont
  {Fujimori}, \citenamefont {Shen}, \citenamefont {Ishikado},\ and\
  \citenamefont {Uchida}}]{Yamasaki2007}%
  \BibitemOpen
  \bibfield  {author} {\bibinfo {author} {\bibfnamefont {T.}~\bibnamefont
  {Yamasaki}}, \bibinfo {author} {\bibfnamefont {K.}~\bibnamefont {Yamazaki}},
  \bibinfo {author} {\bibfnamefont {A.}~\bibnamefont {Ino}}, \bibinfo {author}
  {\bibfnamefont {M.}~\bibnamefont {Arita}}, \bibinfo {author} {\bibfnamefont
  {H.}~\bibnamefont {Namatame}}, \bibinfo {author} {\bibfnamefont
  {M.}~\bibnamefont {Taniguchi}}, \bibinfo {author} {\bibfnamefont
  {A.}~\bibnamefont {Fujimori}}, \bibinfo {author} {\bibfnamefont {Z.-X.}\
  \bibnamefont {Shen}}, \bibinfo {author} {\bibfnamefont {M.}~\bibnamefont
  {Ishikado}}, \ and\ \bibinfo {author} {\bibfnamefont {S.}~\bibnamefont
  {Uchida}},\ }\href {\doibase 10.1103/PhysRevB.75.140513} {\bibfield
  {journal} {\bibinfo  {journal} {Phys. Rev. B}\ }\textbf {\bibinfo {volume}
  {75}},\ \bibinfo {pages} {140513} (\bibinfo {year} {2007})}\BibitemShut
  {NoStop}%
\bibitem [{\citenamefont {Perfetti}\ \emph {et~al.}(2007)\citenamefont
  {Perfetti}, \citenamefont {Loukakos}, \citenamefont {Lisowski}, \citenamefont
  {Bovensiepen}, \citenamefont {Eisaki},\ and\ \citenamefont
  {Wolf}}]{Perfetti2007}%
  \BibitemOpen
  \bibfield  {author} {\bibinfo {author} {\bibfnamefont {L.}~\bibnamefont
  {Perfetti}}, \bibinfo {author} {\bibfnamefont {P.~A.}\ \bibnamefont
  {Loukakos}}, \bibinfo {author} {\bibfnamefont {M.}~\bibnamefont {Lisowski}},
  \bibinfo {author} {\bibfnamefont {U.}~\bibnamefont {Bovensiepen}}, \bibinfo
  {author} {\bibfnamefont {H.}~\bibnamefont {Eisaki}}, \ and\ \bibinfo {author}
  {\bibfnamefont {M.}~\bibnamefont {Wolf}},\ }\href {\doibase
  10.1103/PhysRevLett.99.197001} {\bibfield  {journal} {\bibinfo  {journal}
  {Phys. Rev. Lett.}\ }\textbf {\bibinfo {volume} {99}},\ \bibinfo {pages}
  {197001} (\bibinfo {year} {2007})}\BibitemShut {NoStop}%
\end{thebibliography}%

%\begin{thebibliography}{10}
%\expandafter\ifx\csname url\endcsname\relax
%  \def\url#1{\texttt{#1}}\fi
%\expandafter\ifx\csname urlprefix\endcsname\relax\def\urlprefix{URL }\fi
%\providecommand{\bibinfo}[2]{#2}
%\providecommand{\eprint}[2][]{\url{#2}}

%\end{thebibliography}

\end{document}